# Iron isotope anomalies and the origin of the Earth

Timo Hopp[*], Shengyu Tian, and Thorsten Kleine

**Affiliation:** Max Planck Institute for Solar System Research; 37077 Göttingen, Germany.

* Corresponding author. Email: hopp@mps.mpg.de

**Abstract**

Understanding the origin of the Earth requires determining the original formation location of its building material. Based on the similar Fe isotopic composition of Earth's mantle and Ivuna-type (CI) chondrites, a prior study has argued that Earth formed by accretion of sunward-drifting pebbles from the outer Solar System. Here, using new high-precision Fe isotopic data, we show however that CI chondrites and Earth's mantle have distinct Fe isotopic composition when the neutron-rich $^{58}$Fe is also considered. This observation rules out that the Fe in Earth's mantle derives from CI chondrite-like material and demonstrates that Earth did not form by accretion of sunwards-drifting pebbles. We show that the Fe in Earth's mantle instead derives from the inner Solar System, and has been partly or wholly delivered by bodies from the innermost disk that remained unsampled among meteorites. This provenance of terrestrial Fe is consistent with the classical model of Earth's formation by hierarchical growth among inner Solar System planetesimals and planetary embryos.

**1. Introduction**

Deciphering the formation history of the Earth requires knowledge of the dynamical processes by which the Earth accreted most of its mass. In the classic model, this happened by collisions among planetesimals and Moon- to Mars-sized planetary embryos over a timescale of several tens of millions of years (Myr) (Chambers, 2004). By contrast, in the competing pebble accretion model most of the Earth grew during the ~4 Myr lifetime of the gaseous disk by accretion of mm-cm sized, sunward-drifting pebbles which themselves originated in the outer Solar System (Johansen et al., 2021; Onyett et al., 2023; Schiller et al., 2020, 2018). Thus, a key difference between these two models lies in the provenance of the accreted materials (Morbidelli et al., 2025). Whereas in the classic model most of the Earth's mass derives from inner Solar System objects (Burkhardt et al., 2021), the pebble accretion model predicts an overall large contribution (~25-40%) of outer Solar System material and that this material has predominately been added during the later stages of accretion (Schiller et al., 2020, 2018). Consequently, quantifying the contribution of outer Solar System material to Earth, and the timing of its addition, is essential for understanding the dynamical processes that formed Earth (Burkhardt et al., 2021; Schiller et al., 2018).

The original formation locations of Earth's building materials can be investigated using nucleosynthetic isotope anomalies. These arise through the heterogeneous distribution of presolar material in the solar protoplanetary disk and allow distinguishing between non-carbonaceous (NC) and carbonaceous chondrite (CC) type materials which are presumed to represent the inner and outer Solar System, respectively (Budde et al., 2016; Warren, 2011). Thus, the comparison of the isotopic composition of the bulk silicate Earth (BSE) to that of NC and CC meteorites allows determining the fraction of outer Solar System accreted by the Earth. Importantly, the CC fraction determined for different elements may vary, depending on whether an element is lithophile ('silicate-loving') or siderophile ('iron-loving') (Dauphas, 2017). For the former, the BSE's isotopic composition provides an integrated signature of Earth's bulk accreted materials (i.e., the bulk CC fraction in Earth). By contrast, siderophile elements record only later stages of accretion because those from earlier-accreted objects have been removed to Earth's core and Earth's mantle is unlikely to have equilibrated completely with the bulk core during late-stage accretionary events. Thus, by comparing the CC fractions recorded in lithophile and siderophile elements in the BSE it is possible to reconstruct the evolution of CC accretion to Earth (Budde et al., 2019; Dauphas et al., 2024; Nimmo et al., 2024; Schiller et al., 2020). This in turn makes siderophile elements particularly useful for distinguishing between the classic and pebble accretion models for Earth's formation, because for pebble accretion Earth's late growth stages should be dominated by CC material (Kleine and Nimmo, 2025; Morbidelli et al., 2025; Schiller et al., 2020, 2018).

Of the siderophile elements, Fe is particularly important because its isotopic composition in the BSE likely records a large fraction of the material added during the second half of Earth's accretion (Dauphas, 2017; Hopp et al., 2022b; Schiller et al., 2020). Moreover, Fe becomes less siderophile under more oxidizing conditions, and so the addition of presumably more oxidized CC material is expected to have left a significant imprint on the BSE's Fe isotopic composition. Thus, because the pebble accretion model predicts that the late stages of Earth's accretion are dominated by volatile-rich and oxidized CC pebbles, the BSE is expected to have a CC-like Fe isotopic composition.

Iron has four stable isotopes, $^{54}$Fe (5.85%), $^{56}$Fe (91.75%), $^{57}$Fe (2.12%), and the low abundant $^{58}$Fe (0.28%), which are predominantly produced in massive stars during core collapse and type Ia supernovae (Heger et al., 2014). Previous studies have shown that most meteorites display nucleosynthetic variations for measurements of the three isotopes $^{54}$Fe, $^{56}$Fe, and $^{57}$Fe (Hopp et al., 2022b; Schiller et al., 2020; Spitzer et al., 2025). Moreover, the BSE and Ivuna-type (CI) chondrites [and samples of asteroid Ryugu which have a CI chondrite-like isotopic composition; (Yokoyama et al., 2023)] have strikingly similar $^{54}$Fe isotopic compositions (Hopp et al., 2022a; Schiller et al., 2020). Given that CI chondrites are thought to represent the composition of a large fraction of the outer disk and, hence, the likely composition of pebbles drifting towards the Sun, the CI chondrite-like $^{54}$Fe isotopic composition of the BSE was taken as evidence for pebble accretion origin of the Earth (Schiller et al., 2020). However, given the more siderophile character of Ni compared to Fe, the BSE should also have CI chondrite-like Ni isotopic composition, which is not observed (Hopp et al., 2022b).

Given these ambiguities, we have revisited the Fe isotopic composition of CI chondrites and report new high-precision Fe isotopic data for three CI chondrites. Importantly, unlike some prior studies (Schiller et al., 2020), we also report data for the neutron-rich and low-abundant isotope $^{58}$Fe. We will show below that the $^{58}$Fe data are key for understanding the origin of Fe isotopic variations among Solar System materials, for using these variations to determine the provenance of Earth's building materials, and ultimately to understand the dynamical processes that built the Earth.

## 2. Samples and Methods

The samples selected for this study include the three CI chondrites Orgueil, Ivuna, and Oued Chebeika 002, the Vigarano-type (CV) chondrite Allende, and the terrestrial geological reference materials BCR-2 and BHVO-2 (Table 1). Sample powders of the CI chondrites were made from pieces weighing ~1.1 g for Orgueil, ~1.3 g for Oued Chebeika, and ~0.11 g for Ivuna. Aliquots of approximately 3-10 mg of each sample powder were digested and processed for Fe isotopic measurements following established protocols (Dauphas et al., 2004; Hopp et al., 2025) (see Supplementary Materials). The Fe isotopic measurements were conducted on a ThermoFisher Scientific Neoma multicollector inductively coupled plasma mass spectrometer (MC-ICP-MS) at the Max Planck Institute for Solar System Research in Göttingen, following established techniques (Hopp et al., 2025, 2022a, 2022b) (see Supplementary Materials).

The data are internally-normalized to $^{57}$Fe/$^{56}$Fe=0.023095 using the exponential law (Marechal et al., 1999) and are reported as deviations from the IRMM-524A reference standard in parts per million, using the μ-notation: $\mu^{5X}Fe = [(^{5X}Fe/^{56}Fe)^*_{sample}/(^{5X}Fe/^{56}Fe)^*_{IRMM-524A} - 1] \times 10^6$. Note that although we report the data as variations in $\mu^{54}Fe$ and $\mu^{58}Fe$, this does not necessarily mean that the anomalies themselves are on these two isotopes. Instead, the observed $\mu^{58}Fe$ and $\mu^{54}Fe$ variations may also reflect variations in $^{56}Fe$ and/or $^{57}Fe$, i.e. the two isotopes used for correction of natural and instrumental mass fractionation. Importantly, the observed $\mu^{58}Fe$ and $\mu^{54}Fe$ variations cannot be reproduced by anomalies on only one isotope, but require anomalies on at least two of the Fe isotopes, which always include either $^{58}Fe$ or $^{54}Fe$, or both (Fig. S1).

Mass-dependent isotopic fractionation could potentially lead to apparent mas-independent isotopic variations if the mass-dependent fractionation is large and not following the exponential law used for internal normalization (Dauphas and Schauble, 2016; Heard et al., 2020). To assess the magnitude of such effects, we also report mass-dependent Fe isotope data for all samples of this study, obtained by standard-sample bracketing of the same measurements used for determining the Fe isotope anomalies. The mass-dependent Fe isotope data are reported in the δ-notation as $\delta^{5X}Fe = (^{5X}Fe/^{54}Fe)_{sample}/(^{5X}Fe/^{54}Fe)_{IRMM-524A} - 1] \times 10^3$.

## 3. Results

The new Fe isotopic data are reported in Table 1 and shown in Fig. 1. The overall mass-dependent Fe isotopic variations ($\delta^{56}Fe$ and $\delta^{57}Fe$) are small, demonstrating that any potential effect of non-exponential mass-dependent fractionation on the internally normalized isotope ratios (i.e. $\mu^{54}Fe$ and $\mu^{58}Fe$) are negligible.

The two terrestrial samples display no Fe isotope anomalies and their $\mu^{54}Fe$ and $\mu^{58}Fe$ values agree with the composition of the BSE as estimated based on a large set of terrestrial samples (Hopp et al., 2025) (Fig. 1). For the CV chondrite Allende, we obtained $\mu^{54}Fe = +27\pm6$ and $\mu^{58}Fe = +4\pm8$, consistent with previous analyses of CV chondrites (Fig. 1), including two analyses of different aliquots of the same sample powder (Hopp et al., 2022a; Zhu et al., 2025). The measured Fe isotopic compositions of the three CI chondrites are indistinguishable from each other and define average values of $\mu^{54}Fe = -1\pm2$ (2SD) and $\mu^{58}Fe = +14\pm4$ (2SD). These results are consistent with prior studies showing that CI chondrites and the BSE have identical $\mu^{54}Fe$ (Hopp et al., 2022a; Schiller et al., 2020). The resolvable $\mu^{58}Fe$ excess agrees with previously reported data for CI chondrites and Ryugu samples, which also show small $\mu^{58}$Fe excesses, albeit with more scatter (Fig. 1).

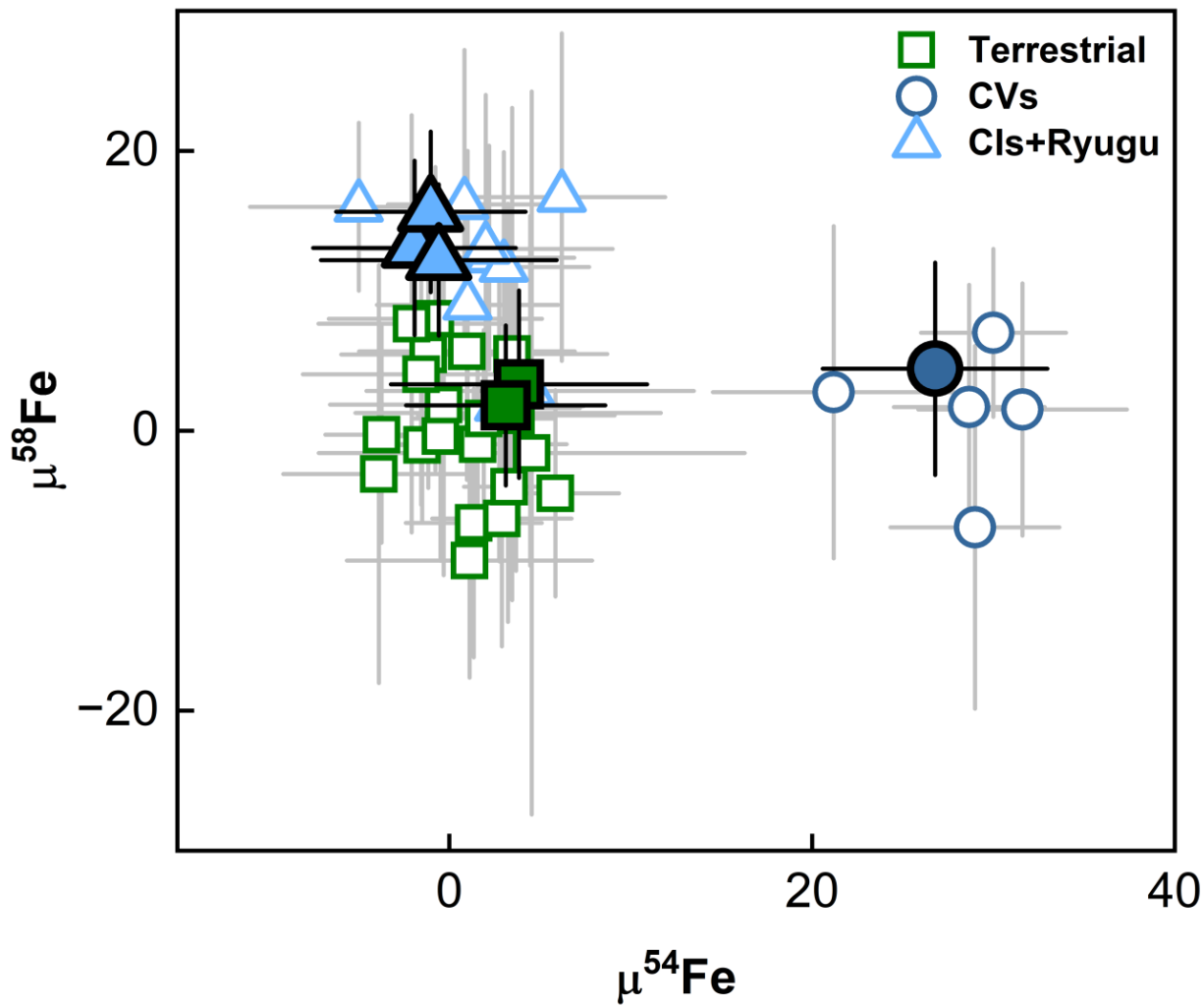

**Fig. 1.** Iron isotopic compositions of CI chondrites, CV chondrites, and terrestrial rocks. The $\mu^{54}$Fe and $\mu^{58}$Fe data for different sample types (see legend) from this study (solid symbols) together with data from previous works (open symbols) (Table S1). Error bars indicate 95% confidence intervals. CV chondrites have similar $\mu^{58}$Fe but elevated $\mu^{54}$Fe compared to terrestrial samples. CI chondrites have similar $\mu^{54}$Fe but elevated $\mu^{58}$Fe compared to terrestrial samples.

**Table 1.** Mass-independent ($\mu^{54}Fe$ and $\mu^{58}Fe$) and mass-dependent ($\delta^{56}Fe$ and $\delta^{57}Fe$) Fe isotopic compositions of CI chondrites, CV chondrite Allende, and geological reference materials. Uncertainties of individual samples are 95% confidence intervals of the mean of *N* analyses. For definition of μ- and δ-values see text.

| Sample | N | $\mu^{54}Fe$ | $\mu^{58}Fe$ | $\delta^{56}Fe$ | $\delta^{57}Fe$ |
|---|---|---|---|---|---|
| *Terrestrial* | | | | | |
| BCR-2 | 45 | +4±7 | +3±7 | +0.04±0.02 | +0.05±0.03 |
| BHVO-2 | 45 | +3±5 | +2±6 | +0.11±0.01 | +0.16±0.01 |
| *CV chondrites* | | | | | |
| Allende MS-A | 30 | +27±6 | +4±8 | −0.05±0.01 | −0.06±0.02 |
| *CI chondrites* | | | | | |
| Orgueil | 45 | −2±6 | +13±6 | −0.05±0.01 | −0.07±0.01 |
| Ivuna | 45 | −1±5 | +16±6 | +0.00±0.01 | −0.01±0.01 |
| Oued Chebeika 002 | 45 | −1±6 | +12±5 | +0.02±0.01 | +0.03±0.01 |
| Average (2SD) | | −1±2 | +14±4 | | |

We performed several statistical tests to assess the significance of the $\mu^{58}Fe$ difference between the BSE and CI chondrites (For details see Supplementary Materials). For these tests we combined our new data with previously published data for terrestrial samples (Hopp et al., 2025), CI chondrites (Gattacceca et al., 2025; Hopp et al., 2022a; Zhu et al., 2025), and samples from asteroid Ryugu (Hopp et al., 2022a; Shollenberger et al., 2025) (Table S1). For some of the tests we also included the CV chondrites, because a large number of analyses exist for this meteorite group (Hopp et al., 2022a; Zhu et al., 2025). A two-sample *t*-test rejects the null hypothesis that CI chondrites and the BSE have the same mean $\mu^{58}Fe$ value at the 95% confidence level ($p \approx 1.5 \times 10^{-7}$). A hierarchical cluster analysis using both $\mu^{54}Fe$ and $\mu^{58}Fe$ for the BSE, CI, and CV chondrites reveals the presence of three clusters, where the distinction between CV chondrites and BSE/CI chondrites is based on $\mu^{54}Fe$, while the distinction between BSE and CI chondrites is based on $\mu^{58}Fe$ (Fig. S2). Finally, a *k*-means clustering analysis also reveals the presence of three statistically significant clusters based on the elbow method (Fig. S3). Additionally, the clustering analysis shows that two individual measurements for CI chondrites and two individual measurements of terrestrial samples are outliers because they were assigned to the wrong cluster (Fig. S3B; Table S1). However, including or excluding these data from the calculated average isotopic compositions of CI chondrites/Ryugu and the BSE does not significantly change these compositions: CI chondrites/Ryugu have average values of $\mu^{54}Fe = +1\pm1$ and $\mu^{58}Fe = +12\pm2$, which change to $\mu^{54}Fe = +1\pm2$ and $\mu^{58}Fe = +13\pm2$ when the two outlier are excluded (all uncertainties are 95% confidence intervals) (Table S1). Similarly, the terrestrial samples define average values of $\mu^{54}$Fe = 0±1 and $\mu^{58}$Fe = +1±2, or $\mu^{54}$Fe = +1±1 and $\mu^{58}$Fe = +1±2 when the two outliers are excluded (Table S1). Combined, these data show that while the BSE and CI chondrites have indistinguishable $\mu^{54}$Fe, they are characterized by distinct $\mu^{58}$Fe values (Fig. 1).

## 4. Discussion

### 4.1. Nucleosynthetic Fe isotopic variations among meteorites and planets

The identification of nucleosynthetic $^{58}$Fe anomalies and thus the ability to assess $\mu^{54}$Fe and $\mu^{58}$Fe variations together uncovers a new perspective on the origin of the Fe isotope anomalies (Fig. 2). In particular, the $\mu^{58}$Fe–$\mu^{54}$Fe systematics of meteorites reveal that CI chondrites, CC meteorites, and NC meteorites have systematically different Fe isotopic compositions. Except for CI chondrites, all meteorites have positive $\mu^{54}$Fe values relative to the BSE. Moreover, where the CC meteorites have more positive $\mu^{54}$Fe values than NC meteorites. The CC meteorites also tend to have positive $\mu^{58}$Fe while NC meteorites tend to have negative $\mu^{58}$Fe. The CI chondrites stand out by having the lowest $\mu^{54}$Fe (which is indistinguishable from the BSE) but highest $\mu^{58}$Fe among all meteorites (Fig. 2). For both NC and CC meteorites, the $\mu^{54}$Fe and $\mu^{58}$Fe isotope anomalies are correlated with anomalies in other elements, such as $\mu^{54}$Cr and $\mu^{96}$Zr (Fig. 3). For the carbonaceous chondrites, these correlated isotopic variations for the Fe-group elements $^{54}$Cr and $^{50}$Ti have been attributed to variable mixing of CI chondrite-like matrix, chondrules, and refractory inclusions, each of which has distinct isotopic compositions (Hellmann et al., 2023) (Fig. S4A). Of note, the $\mu^{58}$Fe values of CC chondrites and CI chondrites are correlated with their matrix mass fractions (Fig. S4B) as well as with $\mu^{54}$Cr and $\mu^{50}$Ti (Fig. 3A; Fig. S5A), indicating that the $\mu^{58}$Fe variations among CI and CC chondrites mimic isotopic variations in other neutron-rich Fe-peak elements like Ti and Cr. By contrast, while the $\mu^{54}$Fe variations among the CC chondrites also correlate with $\mu^{54}$Cr, $\mu^{50}$Ti, and the matrix mass fractions, CI chondrites plot off these trends (Fig. 3B; S4C; S5B), highlighting the uniquely distinct $^{54}$Fe isotopic composition of the CI chondrites.

Similar isotopic systematics can be observed for Ni, where the $\mu^{64}$Ni variations among the carbonaceous chondrites are correlated with $^{54}$Cr and $^{50}$Ti isotope anomalies, while for $\mu^{60}$Ni the CI chondrites plot off these correlations (Spitzer et al., 2024). Thus, in carbonaceous chondrites the isotope anomalies involving neutron-rich isotopes $^{58}$Fe and $^{64}$Ni exhibit the same systematic behavior as other Fe-group elements, while for anomalies involving the neutron-poor isotopes $^{54}$Fe and $^{60}$Ni the CI chondrites deviate from this systematic behavior. This requires a distinct carrier for $^{54}$Fe and $^{60}$Ni in the carbonaceous chondrites, which must be poor in $^{58}$Fe and $^{64}$Ni to not disturb the systematic behavior of these isotopes among all carbonaceous chondrites, including the CI chondrites. Spitzer et al. (2024) suggested that this carrier may be isotopically anomalous Fe-Ni metal grains having an isotopic composition similar to some refractory inclusions showing negative µ$^{54}$Fe (Shollenberger et al., 2019), and which may be related to amoeboid olivine aggregates. Regardless of the origin and nature of this carrier, the finding that variations in µ$^{54}$Fe (and µ$^{60}$Ni) and µ$^{58}$Fe (and µ$^{64}$Ni) reflect the heterogeneous distribution of different carriers having a distinct nucleosynthetic heritage has important implications for utilizing these isotope anomalies to trace the origin of Earth's building materials (see below).

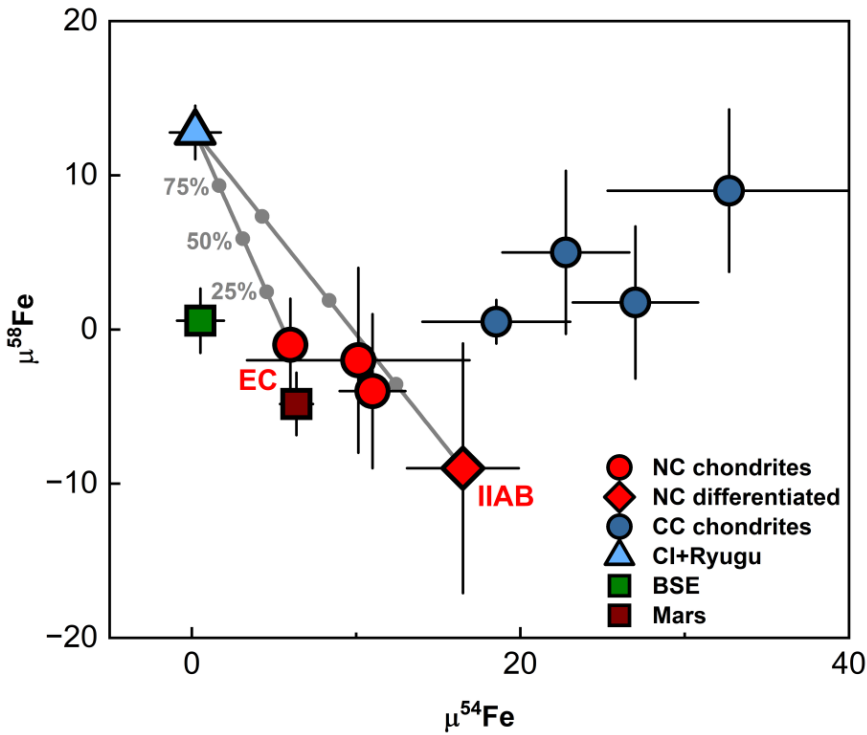

**Fig. 2.** Iron isotopic compositions of Earth's mantle, Mars, and meteorites. The average µ$^{54}$Fe and µ$^{58}$Fe for different sample types (see legend) calculated from data of this study and previous works (Table S2; Data S1). Error bars indicate 95% confidence intervals. CI chondrites have similar µ$^{54}$Fe but elevated µ$^{58}$Fe compared to Earth's mantle (BSE). CC chondrites display more positive µ$^{54}$Fe and µ$^{58}$Fe values than NC meteorites. Mars has a similar µ$^{54}$Fe as NC chondrites but negative µ$^{58}$Fe resolved from the BSE. IIAB iron meteorites display the largest Fe isotope anomalies in the NC reservoir (Table S2). The grey lines are mixing lines between CI chondrites and enstatite chondrites (EC), or between CI chondrites and differentiated NC meteorites (IIAB irons). Tick marks indicate the fraction of CI material in the mixture. No mixture of CI chondrite-like material with any meteorite group can produce the composition of the BSE.

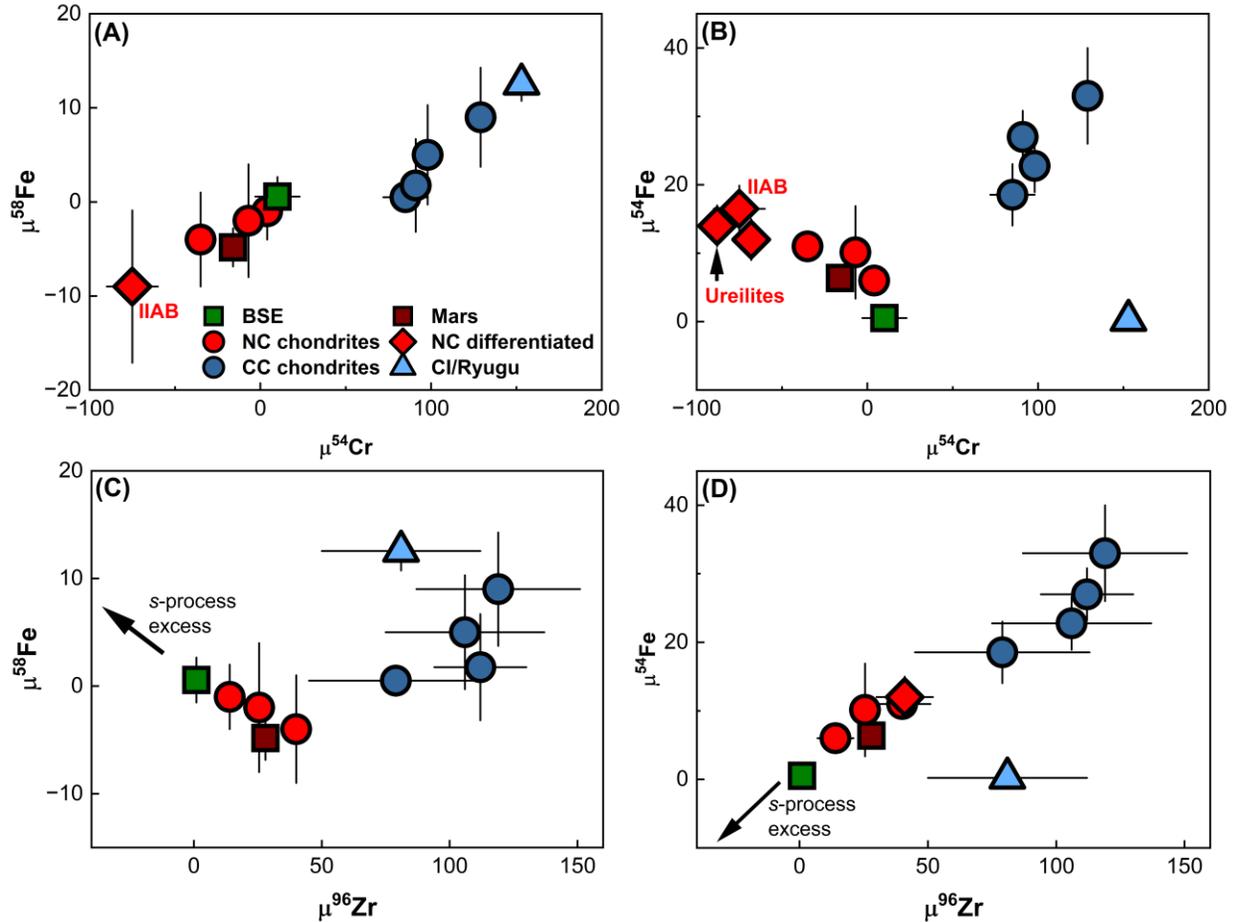

**Fig. 3.** Isotopic anomalies of Fe, Cr, and Zr in Earth's mantle (BSE), Mars, and meteorites. The average isotopic compositions of different sample types (see legend) were calculated based on data from this study and previous works (Table S2; Data S1). Error bars indicate 95% confidence intervals. **(A)** In $\mu^{58}$Fe versus $\mu^{54}$Cr space CI chondrites plot furthest away from the BSE, and display correlated isotope anomalies with other CC chondrites. The BSE, Mars, and NC meteorites also define a correlation, which is offset from the CC correlation towards more negative values. **(B)** In $\mu^{54}$Fe versus $\mu^{54}$Cr space CI chondrites are distinct from the BSE and all other meteorites. In **(C)** $\mu^{58}$Fe versus $\mu^{96}$Zr and **(D)** $\mu^{54}$Fe versus $\mu^{96}$Zr space the BSE plots on the extension of the NC trend pointing towards an *s*-process enriched composition and away from CC meteorites and CI chondrites.

*4.2. Implications for the accretion of the terrestrial planets*

The similarity of CI chondrites and the BSE in $\mu^{54}$Fe has been used to argue that all of the Fe in Earth's mantle was delivered by CI chondrite-like material (Schiller et al., 2020). Since the Fe in the present-day BSE reflects the average isotopic composition of the materials delivered during approximately the second half of Earth's growth (see above), and because CI chondrites represent the presumed composition of pebbles from the outer disk, this observation has been taken as strong evidence for formation of the Earth by pebble accretion (Schiller et al., 2020). However, our results reveal that CI chondrites and the BSE have distinct $\mu^{58}$Fe, demonstrating that the BSE's Fe cannot derive from CI chondrite-like material.

A key observation from the new Fe isotopic data is that the Fe isotopic composition of the BSE cannot be reproduced by any combination of known meteorites (Fig. 2). This is because all meteorites have positive $\mu^{54}$Fe values compared to the BSE, which overlaps only with the $\mu^{54}$Fe of CI chondrites; however, as noted above, these cannot be the source of the BSE's Fe because they have positive $\mu^{58}$Fe compared to the BSE. Consequently, at least some of the BSE's Fe must come from objects that remained unsampled among meteorites (Fig. 2). This situation is similar to observations from the isotope systematics of heavier elements (e.g., Mo, Zr), which show that the Earth incorporated an unsampled component whose isotopic composition is characterized by an enrichment in nuclides produced in the slow neutron capture process (*s*-process) of stellar nucleosynthesis (Akram et al., 2015; Burkhardt et al., 2021, 2011; Hopp et al., 2025; Render et al., 2022).

In plots of $\mu^{96}$Zr versus $\mu^{58}$Fe and $\mu^{54}$Fe, the BSE plots on the extension of the trend defined by NC meteorites, towards lower $\mu^{96}$Zr values (i.e. more *s*-process-enriched signatures) (Fig. 3C,D). This observation has two important implications. First, the *s*-enriched, unsampled material accreted by the Earth must be characterized by negative $\mu^{54}$Fe and positive $\mu^{58}$Fe values. As such, this material can account for the offset of the BSE from NC meteorites in the $\mu^{54}$Fe–$\mu^{58}$Fe plot (Fig. 2). Within this framework, the $\mu^{54}$Fe similarity of the BSE and CI chondrites likely arises from two processes: (*i*) mixing between known NC materials having positive $\mu^{54}$Fe with the unsampled component having negative $\mu^{54}$Fe produced the Fe isotopic composition of the BSE; and (*ii*) fractionation of a $^{54}$Fe-poor carrier among the carbonaceous chondrites produced the lower $\mu^{54}$Fe of CI compared to CC chondrites. These two processes are not related to each other, and so the resulting overlap in the $\mu^{54}$Fe values of the BSE and CI chondrites must be considered purely coincidental.

The second implication of the NC trends in $\mu^{96}$Zr–$\mu^{54}$Fe and $\mu^{96}$Zr–$\mu^{58}$Fe is that the Fe in Earth's mantle almost exclusively derives from NC bodies, because otherwise the BSE would plot away from these trends, towards the composition of CI chondrites or CC meteorites. This, however, is not observed. Importantly, an NC origin of the BSE's Fe effectively rules out that the Earth formed by accretion of sunwards-drifting pebbles from the outer Solar System. As shown by Schiller et al. (2020) , this growth mechanism would require that the BSE's Fe was almost entirely delivered by pebbles. Thus, if the Earth were to form by pebble accretion, it would have to be from NC pebbles. This, however, appears unlikely for two reasons. First, pebble accretion inside of the water ice line (*i.e.,* in the NC reservoir) is inefficient (Batygin and Morbidelli, 2022; Morbidelli et al., 2025). Second, to be accreted by the proto-Earth, the NC pebbles would have to derive from a greater heliocentric distance than proto-Earth itself, *i.e.* from a more distant region of the NC reservoir. However, the position of the BSE on the extension of the $\mu^{96}Zr$-$\mu^{54}Fe$ and $\mu^{96}Zr$-$\mu^{58}Fe$ trends defined by NC meteorites indicates that the BSE's Fe predominantly derives from the *s*-enriched unsampled component (Fig. 3C,D), which is thought to derive from a more proximal region of the disk, inside the Earth's orbit (Burkhardt et al., 2021; Hopp et al., 2025).

Together, the new Fe isotopic data indicate that pebble accretion played only a minor, if any, role in Earth's formation. By contrast, the NC origin of the BSE's Fe demonstrates that Earth predominantly accreted from 'local' material in the terrestrial planet region with only minor contributions from outer Solar System objects (Burkhardt et al., 2021; Nimmo et al., 2024; Steller et al., 2022). As such, the new Fe isotopic are consistent with the classic model of oligarchic growth from predominantly inner Solar System planetesimals and planetary embryos (e.g., Morbidelli et al., 2025).

## 5. Conclusions

New high-precision Fe isotopic data reveal that CI chondrites and the BSE have distinct Fe isotopic compositions for isotope ratios involving the previously not considered neutron-rich isotope $^{58}Fe$. Thus, contrary to a prior proposition, the BSE's Fe cannot derive from CI chondrite-like material. Instead, the combined $\mu^{54}Fe$-$\mu^{58}Fe$ isotope systematics reveal that the BSE's Fe was delivered almost exclusively from inner Solar System objects, including bodies that remained unsampled among meteorites and likely originated from the innermost disk. These observations effectively rule out that Earth accreted any significant mass by accretion of sunward-drifting pebbles, but are fully consistent with formation of the Earth by stochastic collisional growth from planetesimals and planetary embryos from the inner Solar System.

# References

Akram, W., Schönbächler, M., Bisterzo, S., Gallino, R., 2015. Zirconium isotope evidence for the heterogeneous distribution of s-process materials in the solar system. Geochim. Cosmochim. Acta 165, 484–500.

Batygin, K., Morbidelli, A., 2022. Self-consistent model for dust-gas coupling in protoplanetary disks. Astron. Astrophys. 666, A19.

Budde, G., Burkhardt, C., Brennecka, G.A., Fischer-Gödde, M., Kruijer, T.S., Kleine, T., 2016. Molybdenum isotopic evidence for the origin of chondrules and a distinct genetic heritage of carbonaceous and non-carbonaceous meteorites. Earth Planet. Sci. Lett. 454, 293–303.

Budde, G., Burkhardt, C., Kleine, T., 2019. Molybdenum isotopic evidence for the late accretion of outer Solar System material to Earth. Nat. Astron. 3, 736–741.

Burkhardt, C., Kleine, T., Oberli, F., Pack, A., Bourdon, B., Wieler, R., 2011. Molybdenum isotope anomalies in meteorites: Constraints on solar nebula evolution and origin of the Earth. Earth Planet. Sci. Lett. 312, 390–400.

Burkhardt, C., Spitzer, F., Morbidelli, A., Budde, G., Render, J.H., Kruijer, T.S., Kleine, T., 2021. Terrestrial planet formation from lost inner solar system material. Sci. Adv 7, 7601.

Chambers, J.E., 2004. Planetary accretion in the inner Solar System. Earth Planet. Sci. Lett. 223, 241–252.

Dauphas, N., 2017. The isotopic nature of the Earth's accreting material through time. Nature 541, 521–524.

Dauphas, N., Hopp, T., Nesvorný, D., 2024. Bayesian inference on the isotopic building blocks of Mars and Earth. Icarus 409, 115805.

Dauphas, N., Janney, P.E., Mendybaev, R.A., Wadhwa, M., Richter, F.M., Davis, A.M., Van Zuilen, M., Hines, R., Foley, C.N., 2004. Chromatographic separation and multicollection-ICPMS analysis of iron. Investigating mass-dependent and -independent isotope effects. Anal. Chem. 76, 5855–5863.

Dauphas, N., Schauble, E.A., 2016. Mass fractionation laws, mass-independent effects, and isotopic anomalies. Annu. Rev. Earth Planet. Sci. 44, 709–783.

Gattacceca, J., Gounelle, M., Devouard, B., Barrat, J.A., Bonal, L., King, A.J., Maurel, C., Beck, P., Roskosz, M., Viennet, J.C., Mukherjee, D., Dauphas, N., Heck, P.R., Yokoyama, T., López García, K., Poch, O., Grauby, O., Harrison, C.S., Vinogradoff, V., Vernazza, P., Tikoo, S., Vidal, V., Rochette, P., AuYang, D., Borschneck, D., Juraszek, J., Clark, B., 2025. Oued Chebeika 002: A new CI1 meteorite linked to outer solar system bodies. Meteorit. Planet. Sci. 60, 1441–1479.

Heard, A.W., Dauphas, N., Guilbaud, R., Rouxel, O.J., Butler, I.B., Nie, N.X., Bekker, A., 2020. Triple iron isotope constraints on the role of ocean iron sinks in early atmospheric oxygenation. Science. 370, 446–449.

Heger, A., Fröhlich, C., Truran, J., 2014. Origin of the elements, in: Treatise on Geochemistry. Elsevier Ltd., pp. 1–14.

Hellmann, J.L., Schneider, J.M., Wölfer, E., Drążkowska, J., Jansen, C.A., Hopp, T., Burkhardt, C., Kleine, T., 2023. Origin of Isotopic Diversity among Carbonaceous Chondrites. Astrophys. J. Lett. 946, L34.

Hopp, T., Dauphas, N., Boyet, M., Jacobson, S., Kleine, T., 2025. The Moon-forming impactor Theia originated from the inner Solar System. Science. 390, 819–823.

Hopp, T., Dauphas, N., et al., 2022a. Ryugu's nucleosynthetic heritage from the outskirts of the Solar System. Sci. Adv. 8, eadd8141.

Hopp, T., Dauphas, N., Spitzer, F., Burkhardt, C., Kleine, T., 2022b. Earth's accretion inferred from iron isotopic anomalies of supernova nuclear statistical equilibrium origin. Earth Planet. Sci. Lett. 577, 117245.

Johansen, A., Ronnet, T., Bizzarro, M., Schiller, M., Lambrechts, M., Nordlund, Å., Lammer, H., 2021. A pebble accretion model for the formation of the terrestrial planets in the solar system. Sci. Adv 7, 444–461.

Kleine, T., Nimmo, F., 2025. Origin of the Earth, in: Anbar, A., Weis, D. (Eds.), Treatise on Geochemistry. pp. 325–381.

Marechal, C.N., Telouk, P., Albarédé, F., Albarédé, A., 1999. Precise analysis of copper and zinc isotopic compositions by plasma-source mass spectrometry. Chem. Geol. 156, 251–273.

Morbidelli, A., Kleine, T., Nimmo, F., 2025. Did the terrestrial planets of the solar system form by pebble accretion? Earth Planet. Sci. Lett. 650, 119120.

Nimmo, F., Kleine, T., Morbidelli, A., Nesvorny, D., 2024. Mechanisms and timing of carbonaceous chondrite delivery to the Earth. Earth Planet. Sci. Lett. 648, 119112.

Onyett, I.J., Schiller, M., Makhatadze, G. V., Deng, Z., Johansen, A., Bizzarro, M., 2023. Silicon isotope constraints on terrestrial planet accretion. Nature 619, 539–544.

Render, J., Brennecka, G.A., Burkhardt, C., Kleine, T., 2022. Solar System evolution and terrestrial planet accretion determined by Zr isotopic signatures of meteorites. Earth Planet. Sci. Lett. 595, 117748.

Schiller, M., Bizzarro, M., Fernandes, V.A., 2018. Isotopic evolution of the protoplanetary disk and the building blocks of Earth and the Moon. Nature 555, 501–510.

Schiller, M., Bizzarro, M., Siebert, J., 2020. Iron isotope evidence for very rapid accretion and differentiation of the proto-Earth. Sci. Adv 6, eaay7604.

Shollenberger, Q.R., Render, J., Wimpenny, J., Armytage, R.M.G., Gunawardena, N., Rolison, J.M., Simon, J.I., Brennecka, G.A., 2025. Elemental and isotopic signatures of Asteroid Ryugu support three early Solar System reservoirs. Earth Planet. Sci. Lett. 664.

Shollenberger, Q.R., Wittke, A., Render, J., Mane, P., Schuth, S., Weyer, S., Gussone, N., Wadhwa, M., Brennecka, G.A., 2019. Combined mass-dependent and nucleosynthetic isotope variations in refractory inclusions and their mineral separates to determine their original Fe isotope compositions. Geochim. Cosmochim. Acta 263, 215–234.

Spitzer, F., Hopp, T., Burkhardt, C., Dauphas, N., Kleine, T., 2025. The evolution of planetesimal reservoirs revealed by Fe-Ni isotope anomalies in differentiated meteorites. Earth Planet. Sci. Lett. 667.

Spitzer, F., Kleine, T., Burkhardt, C., Hopp, T., Yokoyama, T., Abe, Y., Aléon, J., O, C.M., Alexander, D., Amari, S., Amelin, Y., Bajo, K., Bizzarro, M., Bouvier, A., Carlson, R.W., Chaussidon, M., Choi, B.-G., Dauphas, N., Davis, A.M., Di Rocco, T., Fujiya, W., Fukai, R., Gautam, I., Haba, M.K., Tachibana, S., Yurimoto, H., 2024. The Ni isotopic composition of Ryugu reveals a common accretion region for carbonaceous chondrites. Sci. Adv. 10, eadp2426.

Steller, T., Burkhardt, C., Yang, C., Kleine, T., 2022. Nucleosynthetic zinc isotope anomalies reveal a dual origin of terrestrial volatiles. Icarus 386, 115171.

Warren, P.H., 2011. Stable-isotopic anomalies and the accretionary assemblage of the Earth and Mars: A subordinate role for carbonaceous chondrites. Earth Planet. Sci. Lett. 311, 93–100.

Yokoyama, T., Nagashima, K., Nakai, I., Young, E.D., Abe, Y., Aléon, J., Alexander, C.M.O., Amari, S., Amelin, Y., Bajo, K., Bizzarro, M., Bouvier, A., Carlson, R.W., Chaussidon, M., Choi, B.-G., Dauphas, N., Davis, A.M., Di Rocco, T., Fujiya, W., Fukai, R., Gautam, I., Haba, M.K., Hibiya, Y., Hidaka, H., Homma, H., Hoppe, P., Huss, G.R., Ichida, K., Iizuka, T., Ireland, T.R., Ishikawa, A., Ito, M., Itoh, S., Kawasaki, N., Kita, N.T., Kitajima, K., Kleine, T., Komatani, S., Krot, A.N., Liu, M.-C., Masuda, Y., McKeegan, K.D., Morita, M., Motomura, K., Moynier, F., Nguyen, A., Nittler, L., Onose, M., Pack, A., Park, C., Piani, L., Qin, L., Russell, S.S., Sakamoto, N., Schönbächler, M., Tafla, L., Tang, H., Terada, K., Terada, Y., Usui, T., Wada, S., Wadhwa, M., Walker, R.J., Yamashita, K., Yin, Q.-Z., Yoneda, S., Yui, H., Zhang, A.-C., Connolly, H.C., Lauretta, D.S., Nakamura, T., Naraoka, H., Noguchi, T., Okazaki, R., Sakamoto, K., Yabuta, H., Abe, M., Arakawa, M., Fujii, A., Hayakawa, M., Hirata, Naoyuki, Hirata, Naru, Honda, R., Honda, C., Hosoda, S., Iijima, Y., Ikeda, H., Ishiguro, M., Ishihara, Y., Iwata, T., Kawahara, K., Kikuchi, S., Kitazato, K., Matsumoto, K., Matsuoka, M., Michikami, T., Mimasu, Y., Miura, A., Morota, T., Nakazawa, S., Namiki, N., Noda, H., Noguchi, R., Ogawa, N., Ogawa, K., Okada, T., Okamoto, C., Ono, G., Ozaki, M., Saiki, T., Sakatani, N., Sawada, H., Senshu, H., Shimaki, Y., Shirai, K., Sugita, S., Takei, Y., Takeuchi, H., Tanaka, S., Tatsumi, E., Terui, F., Tsuda, Y., Tsukizaki, R., Wada, K., Watanabe, S., Yamada, M., Yamada, T., Yamamoto, Y., Yano, H., Yokota, Y., Yoshihara, K., Yoshikawa, M., Yoshikawa, K., Furuya, S., Hatakeda, K., Hayashi, T., Hitomi, Y., Kumagai, K., Miyazaki, A., Nakato, A., Nishimura, M., Soejima, H., Suzuki, A., Yada, T., Yamamoto, D., Yogata, K., Yoshitake, M., Tachibana, S., Yurimoto, H., 2023. Samples returned from the asteroid Ryugu are similar to Ivuna-type carbonaceous meteorites. Science. 379, eabn7850.

Zhu, K., Dai, B., Cao, X., Tian, S., Chen, L., 2025. O-Fe-Ti isotopic evidence for classifying Oued Chebeika 002 as a CI chondrite and its genetic affinities with CY chondrites, Ryugu, and Bennu. Mon. Not. R. Astron. Soc. Lett. 542, L7–L11.

**Acknowledgments:** We thank J. Hellmann for providing a powder aliquot of CI chondrite Ivuna, and two reviewers as well as the editor Paolo Sossi for their constructive comments. **Funding:** T.K. was supported by the European Research Council Advanced Grant no. 101019380. **CRediT authorship contribution:** Timo Hopp**:** Conceptualization, Investigation, Methodology, Validation, Visualization, Writing – original draft. Thorsten Kleine: Funding acquisition, Resources, Writing – review & editing. Shengyu Tian: Writing – review & editing. **Competing interests:** The authors declare that they have no competing interests. **Data and materials availability:** All isotopic data used in this study are provided in the Supplementary Materials and Data S1.